\documentclass[%
 reprint,
 aps, prl
]{revtex4-2}

\usepackage{bm,color}
\usepackage{graphicx}
\usepackage{amsmath, amssymb}
\usepackage{color}
\usepackage{braket}
\usepackage{physics}
\usepackage{comment}

\begin{document}
\title {Predicted DC current induced by propagating wave in gapless Dirac materials}
\author{Keisuke Kitayama}
\affiliation{Department of Physics, University of Tokyo, Hongo, Bunkyo-ku, Tokyo 113-0033, Japan}
\author{Masao Ogata}
\affiliation{Department of Physics, University of Tokyo, Hongo, Bunkyo-ku, Tokyo 113-0033, Japan}
\affiliation{Trans-scale Quantum Science Institute, University of Tokyo, Bunkyo-ku, Tokyo 113-0033, Japan}
\begin{abstract}
In this paper, we show that the application of propagating waves can induce a DC current even in systems with spatial inversion symmetry. We derive the equation for the DC current induced by propagating waves using two methods: perturbation theory and Floquet theory. These two approaches yield consistent results. We then apply the equation to gapless graphene subjected to propagating waves. A nonzero DC current is predicted in graphene with next nearest neighbor hopping terms. Nonperturbative effects arising from a strong wave amplitude are also discussed within the framework of Floquet theory.
\end{abstract}
\maketitle

\gdef\QR{\bm Q \cdot \bm r}

$Introduction$.---
Nonlinear optical responses, such as shift current and high-harmonic generation, have garnered significant attention recently due to their fundamental scientific interest and promising applications in energy harvesting and ultrafast optoelectronics~\cite{von1981theory, sipe2000second, ghimire2011observation, luu2015extreme, liu2017high, yoshikawa2017high}. Among these, the shift current, an intrinsic bulk photovoltaic effect occurring in noncentrosymmetric crystals, is directly linked to the Berry connection and interband coherence of Bloch electrons~\cite{sipe2000second,morimoto2016topological}. This effect has been extensively investigated in polar materials and ferroelectrics~\cite{young2012first,spanier2016power,glass1974high}. Its mechanism involves the coherent transfer of photogenerated charges via the shift of Bloch wave functions, fundamentally differing from conventional p-n junction photovoltaic mechanisms~\cite{sipe2000second}. However, in systems with spatial inversion symmetry, these second-order nonlinear responses are strictly forbidden under spatially uniform optical fields due to symmetry considerations. Consequently, generating DC photocurrent in centrosymmetric materials remains a significant challenge in nonlinear optics, hindering the exploration of photoinduced phenomena in a vast array of intriguing material systems.

To overcome this limitation, recent studies have explored using spatially modulated driving fields, such as propagating electromagnetic waves or phonons. These fields break the symmetry of the light-matter interaction without altering the material's intrinsic symmetry~\cite{oka2024shockwave, PhysRevB.105.125407, PhysRevLett.131.096301, ahmadabadi2023optical, ji2020photocurrent}. These phenomena are closely related to the old problem of acoustoelectricity~\cite{}, which has a renewed interest as a DC current in two-dimensional electron systems induced by surface acoustic waves in substrates~\cite{}. By leveraging the natural spatial phase gradient, $\nabla\phi \propto \bm{k}$ (where $\bm{k}$ is the wavevector), introduced by propagating waves, asymmetric responses can be induced even in inversion-symmetric materials. Notably, recent work demonstrated novel, dynamically tunable band modifications using propagating waves~\cite{oka2024shockwave}. Such spatially inhomogeneous driving effectively introduces a preferred direction into the system, lifting the symmetry constraints that would otherwise prohibit DC current generation and opening new avenues for manipulating material properties without structural modifications.

In this work, we present a theoretical analysis demonstrating a novel mechanism for DC current generation in centrosymmetric materials. This current is driven solely by a propagating electromagnetic wave with a finite wavevector ($\mathbf{k} \neq 0$). Our mechanism is fundamentally distinct from the conventional shift current, which requires breaking the material's inversion symmetry and is forbidden in the $\mathbf{k}=0$ (uniform field) limit. Furthermore, while prior work utilizing finite-$\mathbf{k}$ fields, such as Ref.~\cite{oka2024shockwave}, demonstrated dynamically tunable band modifications, our study unveils a direct route to DC current generation that was not previously established at finite frequency. Utilizing Floquet theory and symmetry analysis, we identify the microscopic origin of this current, revealing its dependence on the wavevector-dependent nature of the light-matter coupling.

As a concrete and compelling demonstration of this general mechanism, we apply our theory to gapless, centrosymmetric Dirac materials. These materials, including graphene, present intriguing opportunities for photoinduced phenomena due to their unique linear dispersion~\cite{oka2009photovoltaic, PhysRevLett.107.216601,PhysRevB.90.115423,kitayama2020predicted,kitayama2021floquet,kitayama2021predicted,kitayama2022predicted,PhysRevB.110.045127}. However, generating DC photocurrents in these systems has remained a significant challenge due to their inherent symmetry. Proposals to overcome this, for instance by applying  [11], rely on perturbations that fundamentally alter the material's intrinsic properties. In contrast, our work elucidates the conditions for a novel photocurrent in these symmetric materials without such external perturbations. Our results thus establish a new pathway for current control in symmetric quantum materials, opening possibilities for optoelectronic functionalities in systems previously considered inactive under symmetric driving fields.

\begin{figure}[htb]
\begin{center}
\includegraphics[scale=1.0]{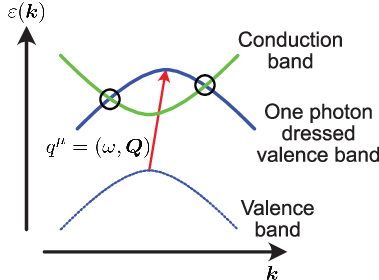}
\caption{Schematic illustration of the Floquet-Bloch two-band model under the influence of a propagating wave. In this scenario, the valence band dressed by one photon is shifted along the space-time vector $(\omega, \bm{Q})$ relative to the original band.}
\label{Fig01}
\end{center}
\end{figure}

{\it Perturbation theory.}---Here, we derive the expression for the DC current induced by a propagating wave using perturbation theory. The time-dependent Hamiltonian for the external field is given by,
\begin{align}
	V(t) &= \sum_i V_0 \, \cos(\bm{Q}\cdot\bm{r}_i - \omega t - \phi) c_i^\dagger c_i \cr
            &= \frac{V_0}{2} \sum_{\bm k} [ 
c_{\bm k + \bm Q}^\dagger c_{\bm k} e^{-i\omega t-i\phi} + 
c_{\bm k - \bm Q}^\dagger c_{\bm k} e^{i\omega t+i\phi}],
\label{eq:potential}
\end{align}
where $c^\dagger_i$ ($c_i$) represents the electron creation (annihilation) operator at site $i$, the trivial spin summation is suppressed,  and $\bm{Q}$ and $\omega$ denote the wave number and frequency of the injected propagating wave, respectively. 
In the second-order response theory, the electric current in the $a$-direction,
$J^a_{\bm q} = e\sum_{\bm k} c_{\bm k-\frac{\bm q}{2}}^\dagger v^a(\bm k)
c^{\phantom{\dagger}}_{\bm k+\frac{\bm q}{2}}$ with momentum $\bm q$
is calculated from the analytic continuation of the response function~\cite{PhysRevB.105.125407}
\begin{align}
&\Phi^{(2)}_{\bm q}(i\omega_{\lambda 1}, i\omega_{\lambda 2}) =
\frac{e V_0^2 k_{\rm B}T}{4} \sum_{n} \sum_{\bm k} {\rm Tr} \biggl[ v^a(\bm k) \cr
&\times \mathcal G \left( \bm k+\frac{\bm q}{2}, i\varepsilon_n \right) 
\mathcal G \left( \bm k +\frac{\bm q}{2} + \bm q_1, i\varepsilon_n - i\omega_{\lambda 1}\right) \cr
& \times \mathcal G \left( \bm k + \frac{\bm q}{2}+ \bm q_1+ \bm q_2, i\varepsilon_n 
-i\omega_{\lambda 1} - i\omega_{\lambda 2}\right) \biggr] \delta_{\bm q+\bm q_1+ \bm q_2, \bm 0},
\label{eq:GeneralFormXX}
\end{align}
where trace (Tr) is taken when the model has a matrix form. 
The analytic continuation is taken for the two cases of
(a) $\bm q_1 = \bm Q, i\omega_{\lambda 1} \rightarrow \hbar \omega + i\delta,
\bm q_2 = -\bm Q, i\omega_{\lambda 2} \rightarrow -\hbar\omega + i\delta$ and 
(b) $\bm q_1 = -\bm Q, i\omega_{\lambda 1} \rightarrow -\hbar\omega + i\delta, 
\bm q_2 = \bm Q, i\omega_{\lambda 2} \rightarrow \hbar\omega + i\delta$, 
%
corresponding to each term in eq. (\ref{eq:potential}). Then, we obtain a DC current with momentum $\bm q=\bm 0$. 
Note that the Matsubara frequency for the external field should satisfy
$\omega_{\lambda 1}>0$ and $\omega_{\lambda 2}>0$ because the perturbation is treated as adiabatic. 
After the summation over $i\varepsilon_n$, the expectaion value of $\langle J^a_{\bm q=\bm 0}\rangle$ becomes
\begin{align}
\langle J^a \rangle &= -\frac{eV_0^2}{4} \sum_{\bm k}
\int \frac{d\varepsilon}{\pi} f(\varepsilon)  {\rm Tr} \biggl[ v^a(\bm k) \cr
&\times \biggl\{ 
G_R(\bm k, \varepsilon) G_R(\bm k + \bm Q, \varepsilon+\hbar\omega)
\left[ {\rm Im} G_R(\bm k, \varepsilon) \right] \cr
&+G_R(\bm k, \varepsilon - \hbar\omega)
\left[ {\rm Im} G_R(\bm k + \bm Q, \varepsilon) \right]
G_A(\bm k, \varepsilon-\hbar\omega)  \cr
&+\left[ {\rm Im} G_R(\bm k, \varepsilon) \right]
G_A(\bm k + \bm Q, \varepsilon+\hbar\omega) G_A(\bm k, \varepsilon)   \cr
&+ ( \bm Q \rightarrow -\bm Q, \omega \rightarrow -\omega) \biggr\} \biggr],
\label{eq:formula}
\end{align}
where $f(\varepsilon)$ is the Fermi distribution function, $f(\varepsilon) = 1/[e^{(\varepsilon-\mu)/k_{\rm B}T}+1]$,
$G_{R(A)}(\bm k, \varepsilon)$ is the retarded (advanced) Green's function, and 
${\rm Im} G_R(\bm k, \varepsilon) = (G_R(\bm k, \varepsilon)-G_A(\bm k, \varepsilon))/2i$. 
In the multi-band case, we use 
$G_{R(A)} (\bm k, \varepsilon) = \sum_j \frac{|u_j(\bm k) \rangle \langle u_j(\bm k) |}{\varepsilon-\varepsilon_j\pm i\Gamma_j}$, 
where $\varepsilon_j = \varepsilon_j(\bm k)$ denotes the $j$-th unpurturbed energy dispersion 
(i.e., without the potential in Eq.~(\ref{eq:potential})). 
If $\bm Q=\bm 0$, we see $\langle J^a \rangle=0$ because the $\bm k$-integral is 
odd reflecting the fact that the current is not induced in the system with the inversion symmetry. 

In this paper, we consider the situation in which the frequency $\omega$ connects two bands, $1$ and $2$ as shown in Fig.~\ref{Fig01}. 
Then, the most dominant contribution among the various matrix elements in Tr in Eq.~(\ref{eq:formula}) is
\begin{align}
\langle J^a \rangle = -\frac{\pi eV_0^2}{4} \sum_{\bm k} 
\left( v_{11}^a(\bm k) - v_{22}^a(\bm k') \right) \frac{
\left| \langle u_1(\bm k) | u_2 (\bm k') \rangle \right|^2}{\Gamma}  \nonumber \\
\left\{ f(\varepsilon_1(\bm k)) - f(\varepsilon_2(\bm k')) \right\} 
\delta(\varepsilon_1(\bm k) - \varepsilon_2(\bm k') + \hbar\omega).
\label{eq:DC_current_Perturbation}
\end{align}
where $\bm k' = \bm k + \bm Q, f_j = f(\varepsilon_j(\bm k))$, and $v_{jj}^a(\bm k) = \partial \varepsilon_j(\bm k)/\hbar\partial k_a$. Here, we have assumed that the relaxation rate $\Gamma$ is small compared with $\varepsilon(\bm k)$, and $k_{\rm B}T$. Then, the $\varepsilon$ integral can be carried out by taking the poles of Green's functions,[reference] which gives a series expansion with respect to $1/\Gamma^n$. Some detailed derivation is given in Supplemental Material.

{\it Floquet theory.}---Next, we derive the expression for the DC current in Flouquet theory, which allows us to map the problem of a periodically driven system onto an effective static problem described by the Floquet Hamiltonian $H_{\rm F}$, defined as $[H_{\rm F}]_{n,m} = H_{n-m} - m\Omega\delta_{n,m}$. Here, $H_n$ denotes the Fourier coefficients of the time-periodic Hamiltonian defined as $H_n = 1/T\int (H_0 + V(t))e^{in\Omega t}$. Using the rotating wave approximation, when the off-diagonal terms of $H_{\rm F}$ are sufficiently smaller than the diagonal terms, $H_{\rm F}$ can be truncated to a 2$\times$2 matrix, given by
\begin{equation}
H_{\rm F} = \left(
	\begin{array}{cc}
		 \varepsilon_1(\bm{k}) + \hbar\Omega &  \frac{V_0}{2} e^{-i\phi}\braket{u_1(\bm{k})}{u_2(\bm{k}^\prime)} \\
		 \frac{V_0}{2} e^{i\phi} \braket{u_2(\bm{k}^\prime)}{u_1(\bm{k})}& \varepsilon_2(\bm{k}^\prime)
	\end{array}
	\right).
	\label{eq:FloquetHam}
\end{equation}
Hereafter, we define $\varepsilon_0$ and $\bm{d}$ such that the Floquet Hamiltonian takes the form $H_{\rm F} = \varepsilon_0 + \bm{d}\cdot\bm{\sigma}$, where $\bm{\sigma}$ is the vector of Pauli matrices.

In the following, we assume that the system is coupled to a fermionic bath with temperature $T$ and dissipation rate $\Gamma$. In this case, using the Floquet-Keldysh formalism, the DC current is given by $J^a = -ie\bm{\mathrm{Tr}}[\tilde{v}^a_{nm}G^<_{mn}]/\hbar$, where $\tilde{v}^a$ is the current operator defined from the Floquet Hamiltonian as $\tilde{v}^a = \partial H_{\rm F}/\partial k^a$, and $G^<_{mn}$ is the lesser Green's function~\cite{morimoto2016topological}. Hereafter, we define $b_0^a$ and $\bm{b}^a$ through $\tilde{v}^a =b_0^a + \bm{b}^a\cdot\bm{\sigma}$. By using $H_{\rm F}$ given in Eq.~(\ref{eq:FloquetHam}), the DC current is expressed as $J^a = j_1^a + j_2^a + j_3^a$ with
\begin{align}
	j_1^a &= \frac{e}{\hbar} \int \frac{d\bm{k}}{(2\pi)^3} \frac{\Gamma/2 (f_1-f_2)}{d^2+\Gamma^2/4}(d_yb_x^a-d_xb_y^a)\label{eq:j1} \\
	j_2^a &= \frac{e}{\hbar} \int \frac{d\bm{k}}{(2\pi)^3}\frac{d_z (f_1-f_2)}{d^2+\Gamma^2/4}(d_xb_x^a+d_yb_y^a)\label{eq:j2}\\
	j_3^a &= -\frac{e}{\hbar} \int \frac{d\bm{k}}{(2\pi)^3}\frac{b_z^a (f_1-f_2)}{d^2+\Gamma^2/4}(d_x^2+d_y^2)\label{eq:j3},
\end{align}
as derived in Ref.~\cite{morimoto2016topological, PhysRevLett.127.126604}. When we consider only the potential given in Eq.~(\ref{eq:potential}), the off-diagonal components of the current operator $\tilde{v}^a$ vanish. As a result, the terms $j_1$ and $j_2$ vanish and the DC current induced by a propagating wave is obtained as,
\begin{align}
	J^a =& \frac{\pi e }{4\hbar} \int \frac{d\bm{k}}{(2\pi)^3} \left( f_2 - f_1\right)\frac{V_0^2|\braket{u_1(\bm{k})}{u_2(\bm{k}^\prime)}|^2}{\sqrt{V_0^2|\braket{u_1(\bm{k})}{u_2(\bm{k}^\prime)}|^2+\Gamma^2}}  \nonumber \\
	&\times \left( v_{11}^a(\bm{k}) - v_{22}^a(\bm{k}^\prime)\right) \delta(\varepsilon_1(\bm{k}) - \varepsilon_2(\bm{k}^\prime)+\hbar \omega).
	\label{eq:DC_current_Floquet}
\end{align}
This expression includes a saturation factor of $1/\sqrt{V_0^2|\braket{u_1(\bm{k})}{u_2(\bm{k}^\prime)}|^2+\Gamma^2}$, which can be approximated as $1/\Gamma$ in the limit of small $V_0$. Therefore, Eq.~(\ref{eq:DC_current_Floquet}) is consistent with the result derived from the perturbation theory in Eq.~(\ref{eq:DC_current_Perturbation}).

Next, we consider the small $\bm{Q}$ cases of Eq.~(\ref{eq:DC_current_Floquet}). In systems with time-reversal symmetry, the contributions up to the second-order in $\bm{Q}$ vanish. The leading-order contribution in $\bm{Q}$ is of the third given by,
\begin{align}
	J^a =& -\frac{\pi e V_0^2}{\hbar\Gamma} \int \frac{d\bm{k}}{(2\pi)^3} |\bm{Q}\cdot\bm{A}_{12}(\bm{k})|^2 \bm{Q}\cdot \frac{\partial v_{22}^a}{\partial \bm{k}} \nonumber \\
	&\times \left( f_2 - f_1\right) \delta(\varepsilon_1(\bm{k}) - \varepsilon_2(\bm{k})+\hbar \omega),
	\label{eq:DC_current_Floquet2}
\end{align}
where the interband Berry connection $\bm{A}_{12}(\bm{k})$ is defined as $\bm{A}_{12}(\bm{k}) = i\mel{u_1(\bm{k})}{\nabla_{\bm{k}} }{u_2(\bm{k})}$. Although this expression depends on the topological quantity $\bm{A}_{12}(\bm{k})$, it appears only in the form of $\bm{Q}\cdot\bm{A}_{12}(\bm{k})$.

{\it Graphene under propagating wave.}---
\begin{figure}[htb]
\begin{center}
\includegraphics[scale=1.0]{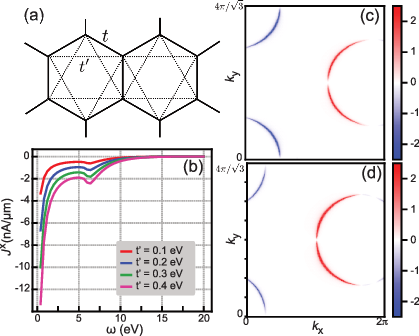}
\caption{(a) Crystal structure of a honeycomb lattice with two types of hopping: nearest-neighbor and next-nearest-neighbor hopping. (b) Calculated DC current as a function of the next-nearest-neighbor hopping amplitude, $t^\prime$. (c,d) $\bm{k}$-space distribution of the integrand in Eq.~(\ref{eq:DC_current_Floquet}) for two different values of $t^\prime$: (c) $t^\prime = 0$~eV and (d) $t^\prime = 1$~eV.}
\label{Fig02}
\end{center}
\end{figure}
As an explicit application of the present theory, we investigate the DC current induced by a propagating wave in gapless graphene. We consider a tight-binding model on the honeycomb lattice, given by
\begin{eqnarray}
	H_0 = t\sum_{\langle i,j \rangle} c^\dagger_i c_j + t^\prime\sum_{\langle\langle i,j \rangle\rangle} c^\dagger_i c_j,
	\label{eq:model}
\end{eqnarray}
as illustrated in Fig.~\ref{Fig02}(a). The parameter $t$ is the nearest-neighbor hopping amplitude, while $t^\prime$ denotes the next-nearest-neighbor (NNN) hopping amplitude.

We examine the dependence of the light-induced DC current on the frequency $\omega$. Figure~\ref{Fig02}(b) shows the calculated DC current as a function of light frequency for different values of the NNN hopping term, $t^\prime$. For this calculation, the parameters are set as follows: the nearest-neighbor hopping $t=3$~eV,  the lattice constant is $a = 0.246$~nm, the wave vector $\bm{Q}=(0.001/a, 0)$, and the amplitude $V_0 = 1$~eV. The induced DC current is found to be approximately proportional to $t^\prime$ and vanishes when $t^\prime = 0$.

This behavior arises from the cancellation of positive and negative peaks in the integrand responsible for the current. As shown in Fig.~\ref{Fig02}(c), in the absence of NNN hopping ($t^\prime = 0$), symmetric semicircle peaks appear at the $\bm{k}$-points satisfying the condition $\varepsilon_1(\bm{k}) - \varepsilon_2(\bm{k}^\prime)+\hbar \omega = 0$. These peaks of opposite sign cancel with each other. In contrast, for a finite NNN hopping term ($t^\prime = 1$~eV, Fig. 2(d)), an imbalance between the positive and negative peaks emerges. This asymmetry leads to a net DC current.

Furthermore, the DC current is enhanced at low frequencies. This enhancement is attributed to the fact that as $\omega$ decreases, the $\bm{k}$-points satisfying the energy condition approach the Dirac points. Proximity to the Dirac points increases the wave function overlap $|\braket{u_1(\bm{k})}{u_2(\bm{k}^\prime)}|$, which effectively amplifies the propagating wave potential $V_0$. Nevertheless, this trend does not continue down to zero frequency. As $\omega$ approaches zero, the $\bm{k}$-points satisfying the condition vanish, causing the induced current to disappear. It is worth noting that this vanishing behavior at low frequencies ($J\to 0$ as $\omega \to 0$) is not observed in Fig.~\ref{Fig02}(b), as our calculations assume a small but finite wave vector $\bm{Q}$.

\begin{figure}[htb]
\begin{center}
\includegraphics[scale=1.0]{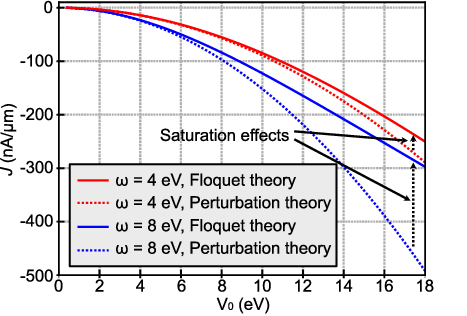}
\caption{ Calculated DC current as a function of wave amplitude $V_0$. The results from Floquet theory (solid lines) and perturbation theory (dashed lines) are compared. The deviation of the Floquet results from the perturbative ones at high amplitudes indicates saturation effects. Each color represents a different frequency $\omega$. The NNN hopping is fixed at $t^\prime = 0.3$ eV.}
\label{Fig03}
\end{center}
\end{figure}

We now turn to the non-perturbative effects that become significant at high wave amplitudes, particularly as the current is enhanced at low frequencies ($\omega$). Figure~\ref{Fig03} shows the calculated DC current as a function of wave amplitude, comparing the results obtained from perturbation theory (dashed lines, Eq.~(\ref{eq:DC_current_Perturbation})) with those from the Floquet theory (solid lines, Eq.~(\ref{eq:DC_current_Floquet})). The deviation of the solid lines from the dashed lines stems from higher-order contributions (third order and above), which constitute the non-perturbative saturation effects. A key finding is that these saturation effects are more pronounced at lower frequencies. While this behavior is analogous to those observed in the shift current~\cite{PhysRevB.110.045127}, the underlying mechanism is distinct. Specifically, in our case, the wave function overlap $|\braket{u_1(\bm{k})}{u_2(\bm{k}^\prime)}|$ plays a crucial role, whereas it is not a key factor in the non-perturbative dynamics of the shift current.

Finally, we examine the dependence of the induced DC current on the dissipation parameter, $\Gamma$. In the perturbative limit (i.e., second order in $V_0$), Eq.~(\ref{eq:DC_current_Perturbation}) predicts that the current is proportional to $1/\Gamma$, implying a divergence as $\Gamma \to 0$. However, nonperturbative effects become dominant when the term $V_0|\braket{u_1(\bm{k})}{u_2(\bm{k}^\prime)}|$ is large compared with $\Gamma$, causing the current to saturate to a constant value. Thus, these nonperturbative effects regularize the current and suppress the unphysical divergence in the limit of small dissipation.



$Summary.$---In summary, we have theoretically predicted that a DC current can be induced by propagating waves even in systems that preserve spatial inversion symmetry. We derived the DC current formula using two complementary approaches—perturbation theory and Floquet theory—and confirmed their consistency. Furthermore, we proposed the honeycomb lattice with next-nearest-neighbor hopping as a promising example of an inversion-symmetric system in which a DC current can be generated by a propagating wave.

$Acknowledgment.$---This work was supported by JSPS KAKENHI (No. 21J20856 and No. 23K03274).
\bibliographystyle{apsrev4-1}
\bibliography{crossresp}
\end{document}